\documentstyle[a4,12pt]{article}

\newcommand{\be}{\begin{equation}}
\newcommand{\ee}{\end{equation}}
\newcommand{\bea}{\begin{eqnarray}}
\newcommand{\eea}{\end{eqnarray}}

\newcommand{\bra}[1]{\left\langle #1 \right|}
\newcommand{\ket}[1]{\left| #1 \right\rangle}
\begin{document}
\vspace*{1cm}
\rightline{BARI-TH/93-163}
\rightline{DSF-T-93/45}
\rightline{INFN-NA-IV-93/45}
\rightline{November 1993}
\vspace*{1cm}
\begin{center}
  \begin{Large}
DEPENDENCE OF THE FORM FACTORS OF $ B \to \pi \ell \nu$
ON THE HEAVY QUARK MASS \\
   \end{Large}
  \vspace{8mm}
  \begin{large}
Pietro  Colangelo \footnote{E-mail address: COLANGELO@BARI.INFN.IT}\\
  \end{large}
  \vspace{6mm}
 Istituto Nazionale di Fisica Nucleare, Sezione di Bari, Italy\\
  \vspace{6mm}
\centerline{and}
  \vspace{3mm}
  \begin{large}
Pietro  Santorelli \footnote{E-mail address: SANTORELLI@NA.INFN.IT}\\
  \end{large}
  \vspace{6mm}
 Istituto Nazionale di Fisica Nucleare, Sezione di Napoli, Italy\\

\end{center}
\begin{quotation}
\vspace*{1.5cm}
\begin{center}
  \begin{bf}
  ABSTRACT
  \end{bf}
\end{center}
\vspace*{0.5cm}

\noindent
We use QCD sum rules to analyze the semileptonic transition
$B \to \pi \ell \nu$ in the limit
$m_b \to \infty$.
We derive the dependence of the form factor $F_1(0)$ on the heavy
quark mass, which is compatible with the expected dependence for the
simple pole model of $F_1(q^2)$.
\end{quotation}

\newpage

Heavy quark and chiral symmetries constrain the semileptonic
decay $B \to \pi \ell \nu$ in the kinematical point where the pion is
at rest in the rest frame of the decaying B meson (zero-recoil point).
As a matter of fact, the form factors which parametrize the hadronic matrix
element governing $B \to \pi \ell \nu$:
\be
\bra{\pi (p^\prime)} V_{\mu} \ket{{ B}(p)}\;=\;F_1 (q^2) \;
(p+p^\prime)_\mu +
\;{m_{B}^2 - m_\pi^2 \over q^2} q_{\mu} \; \lbrack F_0 (q^2) - F_1 (q^2)
\rbrack  \label{zero} \label{eq1}
\ee
\noindent
$(q=p-p^\prime)$ can be written, near the zero-recoil point
$q^2_{max}=(m_B - m_\pi)^2$, as follows
\cite{Wise0,BD,Wolfenstein,Casal,Isgur,Ligeti,PR}
\footnote{ For a review see \cite{Wise}.}:
\be
\left . F_1(q^2)\right |_{q^2 \approx q^2_{max}} = {f_{B^*} \over f_\pi} ~
{g_{B^* B \pi} \over 1 - {q^2/m_{B^*}^2}}  \label{eq2} \ee
\noindent
and
\be
F_0(q^2_{max}) = {f_B \over f_\pi} \hskip 5pt .
\label{eq3} \ee
In eq. (\ref{eq2}) the dominance of the pole of the $B^*$ meson, which is
degenerate with the $B$ meson in the limit $m_b \to \infty$, has been
exploited;
 $f_{B^*}$ and  $f_B$ are  $B^*$ and $B$ meson leptonic constants,
respectively;  $g_{B^* B \pi}$ is the strong $B^* B \pi$ coupling constant.

The phenomenological importance of
eqs. (\ref{eq2},\ref{eq3}) is immediate in the light of the
measurement of $V_{ub}$:
since, neglecting the charged lepton mass,
\be
{ d \Gamma( B \to \pi \ell \nu) \over d q^2} = {G_F^2 \over 24 \pi^3}
|V_{ub}|^2
|F_1(q^2)|^2 |{\vec p}~^\prime_\pi(q^2)|^3 \label{eq4}
\ee
\noindent where
$\vec p~^\prime_\pi(q^2)$ is the pion three-momentum in the B rest frame
at fixed $q^2$,
one could compare the differential rates of
$B \to \pi \ell \nu$ and $D \to \pi \ell \nu$ at the same (small) value of
${\vec p}~^\prime_{\pi}$; the ratio between the rates
\be
\left. { d \Gamma( B \to \pi \ell \nu) / d q^2 \over
 d \Gamma( D \to \pi \ell \nu) / d q^2 } \right|_{same \; \vec p~^\prime_\pi} =
\left. {|V_{ub}|^2 \over |V_{cd}|^2}
{|F_1^{B \to \pi}|^2 \over |F_1^{D \to \pi}|^2}\right|_{same \;
\vec p~^\prime_\pi}
 \label{eq5}
\ee
is given in terms of $f_{B^*} / f_{D^*}$ and
$g_{B^* B \pi} / g_{D^* D \pi}$ which can be measured and/or estimated by
several methods \footnote{We shall not discuss here the role of the breaking
terms either of the chiral symmetry or of the heavy quark flavor symmetry.},
so that a measurement of the left-hand side of eq. (\ref{eq5})
can provide a value of $|V_{ub}|$ with a procedure where the model
dependence is drastically reduced.

This program, proposed in ref.\cite{Ligeti}, finds a relevant
difficulty in the severe phase space suppression
$|{\vec p}~^\prime_\pi(q^2)|^3$ in (\ref{eq4}). For this reason it could be
 useful to investigate
the dependence (if any) of the form factor $F_1(q^2)$ on the mass of the
heavy meson at fixed $q^2$, also far from $q^2_{max}$,
i.e. in kinematical configurations where
not only $m_b$,  but also the momentum of the emitted
pion represent heavy scale parameters.
The aim is  to attempt an extrapolation from
$D \to \pi$ (when accurate experimental data will be available) to
$B \to \pi$.

As we shall show below, such dependence can be predicted
by relativistic QCD sum rules \cite{SVZ} at $q^2=0$.

The form factor $F_1$ in (\ref{eq1}) has been studied by three-point function
QCD sum rules, for a finite value of the
$b$-quark mass, by a number
of authors, adopting an  analysis that can be applied also to
the transitions $D \to (K,\pi) \ell \nu$
 \cite{Aliev, Dosh, Ball, Narison}
\footnote{ For a review see \cite{Col}.}.
By studying the three-point correlator of a pseudoscalar current
having the same quantum numbers of the $B$ meson, of
an axial current interpolating the pion, and of the
flavor-changing current $V_\mu$ in (\ref{eq1}),  the following Borel improved
sum rule can be derived:
\bea
& & f_\pi f_B {m^2_B \over m_b}  F_1(q^2=0)
\exp\left\{- { m^2_B \over M^2} - { m^2_\pi \over M^{\prime 2}} \right\}
\nonumber\\
& = & {1 \over (2 \pi)^2} \int_D ds \; ds^\prime \rho(s, s^\prime) \;
 \exp\left\{- {s \over M^2} - { s^\prime \over M^{\prime 2}} \right\}
\nonumber\\
& - &  {<\bar q q> \over 2} \; \exp\left\{ - {m^2_b \over M^2} \right\} \;
\left[ 1 - {m_0^2 \over 6} \; \left({3 m_b^2  \over 2 M^4 } -
{2 \over M^2}\right) \right] \label{eq6} \eea
\noindent where the integration region $D$ is
$m_b^2 \le s \le s_0, \; 0 \le s^\prime \le min(s^\prime_0, s-m_b^2)$
($s_0, s^\prime_0$ are effective thresholds
separating the resonance region from the continuum); the
perturbative spectral function $\rho$,
 at the lowest order in $\alpha_s$, is given by:
\bea  \rho  (s, s^\prime) & = &
{3 m_b \over 2 (s - s^\prime)^3 } \Big[ 2 {\tilde \Delta} (u - s^\prime) +
s^\prime (u - 4 s)
-  {2 m_b \over (s - s^\prime) }
\Big(  {\tilde \Delta}^2 (u^2 - 3 u s^\prime + 2 s s^\prime)
\nonumber \\ \nonumber \\
& + & 2 {\tilde \Delta} s^\prime (u^2 - 3 u s + 2 s s^\prime)
+ 3 s s^{\prime 2} ( 2 s -u) \Big) \Big] \label{eq7} \eea
\noindent (${\tilde \Delta} =s - m_b^2$, $u=s +s^\prime$);
$M, M^\prime$ are the
Borel parameters associated to $B$ and to the pion channel, respectively;
$<\bar q q>$ is the condensate of dimension 3, whereas
$m_0^2$ is connected to the condensate of dimension 5:
$m_0^2= <\bar q g \sigma G q>/<\bar q q>$.
The mass of the light quarks has been neglected.

The limit $m_b \to \infty$ can be performed by changing the
variables associated
to the heavy quark channel in terms of low-energy variables.
Using:
$s=m_b^2 +2 m_b \omega$,
$s_0=m_b^2 +2 m_b \omega_0$,
$M^2=m_b T $ and $m_B = m_b + \Lambda / 2$,
i.e. using the same procedure adopted in \cite{Neub} in the analysis of
the semileptonic transition
$({\bar q} Q) \;\to  ({\bar q} Q^\prime) \; \ell \nu$,
eq.(\ref{eq6}) can be
written as follows:
\newpage
\bea
& & f_\pi (\sqrt{m_b} f_B) \left(1 + {\Lambda \over m_b}\right)
(\sqrt{m_b} F_1(0))
\exp\left\{- { \Lambda\over T} - { m^2_\pi \over M^{\prime 2}} \right\}
\nonumber\\
& = &
  {3 \over \pi^2 m_b} \int_0^{\omega_0} d\omega \; \int_0^{s^\prime_0}
ds^\prime \omega
 \exp\left\{- {2 \omega \over T} - { s^\prime \over M^{\prime 2}} \right\}
-  {<\bar q q> \over 2} \;
\left[ 1 - {m_0^2 \over 4 T^2}\right] \hskip 5pt .\label{eq8} \eea

{}From this equation the behavior of $F_1(0)$ versus $m_b$ can be
derived. In fact, $\Lambda$, which is related to the binding energy of the
heavy-light quark system, remains finite in the infinite heavy quark mass
limit;
therefore, since the scaling law for $f_B$, for $m_b \to \infty$, is
$ 1/\sqrt{m_b}$ modulo logs, one obtains that
also $F_1(0)$ scales as $1/\sqrt{m_b}$: this is a consequence of the fact
that the perturbative term is subleading in the heavy quark limit,
the sum rule being determined by the non perturbative $D=3$ and $D=5$
contributions.
A similar result has been found for the form
factors $V$ and $A_1$ governing $B \to \rho \ell \nu$ and for the form factor
of the rare $B \to K^* \gamma $ decay \cite{CDNP}.
It is worth observing that the scaling law $1/\sqrt{m_b}$
for $F_1$ at $q^2=0$
is compatible with the correct dependence of $F_1$ on $m_b$ at $q^2_{max}$ in
eq. (\ref{eq2}):
\be
F_1(q^2_{max}) = {f_{B^*} \; m_{B^*} \; g_{B^* B \pi} \over 2 \; f_\pi \;
(\delta_B + m_\pi)}
\simeq {\sqrt{m_b} \over \delta_B +  m_\pi}
\label{eq8a}\ee
\noindent
(with $\delta_B= m_{B^*} - m_B = {\cal O} (1 / m_b)$)
 and with a simple pole evolution from   $q^2_{max}$ to
$q^2=0$.  We shall discuss this point at the end of the paper;
here we want to show that the variables adopted above
are true low energy variables for the system we are considering.

In order to do that, let us analyze the form factors of $B \to \pi \ell \nu$
in the heavy quark effective theory (HQET)
\cite{HQET}. As discussed in ref.\cite{Ligeti},
the matrix element in (\ref{eq1}) can be written, in the framework of HQET,
in terms of the $B$ meson velocity:
$v = p / m_B$,  and of  the energy of the emitted pion in the $B$ rest frame:
\be
v \cdot p^\prime = { m_B^2 + m_\pi^2 -q^2 \over 2 m_B} \hskip 5pt.
\label{eq9}  \ee
\noindent
Accordingly, eq.(\ref{eq1}) can be rewritten as follows:
\be
{\bra{\pi (p^\prime)} V_{\mu} \ket{{ B}(v)} \over \sqrt{m_B} }
\;=\;2 \; f_1 (v \cdot p^\prime) \; v_\mu \; + \; 2 \; f_2 (v \cdot p^\prime)
\; {p^\prime_\mu \over (v \cdot p^\prime)} \hskip 5pt, \label{eq10}
\ee
\noindent
where the functions $f_1$ and $f_2$ are universal, in the sense that
 they become independent of the heavy quark mass $m_b$
  in the limit $m_b \to \infty$, for values of $v \cdot p^\prime$
which do not scale as $m_b$.
It is immediate to derive the relations between the
form factors in (\ref{eq1}) and (\ref{eq10}):
\bea
F_1(q^2) & = & \sqrt{m_B} \left \{ {f_1(v \cdot p^\prime) \over m_B} \;
 + {f_2(v \cdot p^\prime) \over (v \cdot p^\prime) }   \right \} \label{eq11}\\
F_0(q^2) & = & {2 m_B^{3 / 2} \over m^2_B - m^2_\pi } \times \nonumber \\
&\times
&\left \{ \left[ f_1(v \cdot p^\prime) + f_2 (v \cdot p^\prime) \right] -
{ (v \cdot p^\prime) \over m_B} \;
\left[ f_1(v \cdot p^\prime) + {m_\pi^2 \over (v \cdot p^\prime)^2}
f_2(v \cdot p^\prime) \right]  \right \} \; . \nonumber \\ \label{eq12}
\eea

QCD sum rules can be employed
in the evaluation of $f_1$ and $f_2$;
the approach is similar to that used in the determination
 of the Isgur-Wise function \cite{Neub}, with the difference
that, in this case, the effective theory is only applied to $B$-meson channel.
The starting point is the correlator
\be
T_{\mu\nu} (k, p^\prime, v \cdot p^\prime ) = i^2 \int dx \; dy \;
e^{i ( p^\prime \cdot x - k \cdot y)} \;
<0| T \{ j_\nu(x) {\hat V}_\mu(0) {\hat J}^\dagger_5(y) \} |0> \hskip 5pt;
\label{eq12a}\ee
\noindent
the axial current $j_\nu(x)$ interpolates the pion; the other currents
are
${\hat J}_5(y)= {\bar q} (y) i \gamma_5 h_v(y)$  and
${\hat V}_\mu(0)= {\bar q} (0)  \gamma_\mu h_v(0)$,
where $h_v$ is the velocity dependent $b-$quark field in the effective theory,
whose residual "off-shell" momentum is $k$.
The matrix element of
${\hat J}_5$ between the vacuum state and the $B$-meson state defines the
scale-dependent universal leptonic constant ${\hat F}(\mu)$:
\be <0| {\hat J}_5 |B(v)> = {\hat F}(\mu) \hskip 5pt ,\label{eq12b} \ee
\noindent where  $\mu$ is the renormalization scale, and
the connection of
${\hat F}$ to the $B-$meson leptonic constant $f_B$ is given by
\be f_B \sqrt{m_B} = C_1(\mu) {\hat F}(\mu) + O(1/m_b) \hskip 5pt,
\label{eq12c} \ee
\noindent $C_1(\mu)$ being a Wilson coefficient.

By applying to the correlator (\ref{eq12a})
 the usual techniques of the QCD sum rules
method, the following Borel improved rules for
$f_1(v \cdot p^\prime)$ and $f_2(v \cdot p^\prime)$ can be worked out:
\bea
& & 2 \; f_\pi \; {\hat F} \; f_1(v \cdot p^\prime) \;
\exp\left\{- { \Lambda \over T} - { m^2_\pi \over M^{\prime 2}} \right\} \; =
\nonumber\\
& = & {1 \over (2 \pi)^2} \int_0^{\omega_0}  d \omega \;
\int_0^{\hat s^\prime} ds^\prime \;
\rho_1(\omega, s^\prime, v \cdot p^\prime) \;
 \exp\left\{- {\omega \over T} - { s^\prime \over M^{\prime 2}} \right\}
\nonumber\\
& - &  <\bar q q>  \;
\left[ 1 - m_0^2  \; \left({1   \over 4 T^2 } +
{2 \over 3} { (v \cdot p^\prime) \over T M^{\prime 2} }\right) \right]
\label{eq13} \eea
\noindent and
\bea
& & {2 \; f_\pi \; {\hat F} \over (v \cdot p^\prime)} \; f_2(v \cdot p^\prime)
\; \exp\left\{- { \Lambda \over T} - { m^2_\pi \over M^{\prime 2}} \right\}
\; = \nonumber\\
& = & {1 \over (2 \pi)^2} \int_0^{\omega_0} d\omega \;
\int_0^{\hat s^\prime} ds^\prime \;
\rho_2(\omega, s^\prime, v \cdot p^\prime) \;
 \exp\left\{- {\omega \over T} - { s^\prime \over M^{\prime 2}} \right\}
\nonumber\\
& - & { m_0^2  \; <\bar q q>    \over 3 T M^{\prime 2}} \hskip 5pt ,
\label{eq14} \eea
\noindent
with the spectral functions $\rho_1$ and $\rho_2$ given by:
\be
\rho_1(\omega, s^\prime, v \cdot p^\prime) \; =
- {3 \over 8} \; {s^\prime \over \Delta^{5\over2}}
\left\{ 10 \omega (v \cdot p^\prime)^2 + 2 \omega s^\prime
-4 (v \cdot p^\prime)^3 - 8 (v \cdot p^\prime) s^\prime
- 3 \omega^2  (v \cdot p^\prime) \right \} \label{eq15}
\ee
\noindent and
\bea
\rho_2(\omega, s^\prime, v \cdot p^\prime) \; & = &
- {3 \over 4} \; {s^\prime \over \Delta^{5\over2}}
\big \{ 8 s^\prime (v \cdot p^\prime)^2 + 4  s^{\prime 2}
-4 \omega (v \cdot p^\prime)^3 \nonumber \\
& - & 8 \omega (v \cdot p^\prime) s^\prime
+ 2 \omega^2  (v \cdot p^\prime)^2
+  \omega^2 s^\prime
\big \} \hskip 5pt; \label{eq16}
\eea
\noindent
 $\omega_0$ is the effective threshold in the heavy quark channel and
the upper integration limit in $s^\prime$ in (\ref{eq13},\ref{eq14}) is
${\hat s^\prime} = min ( s^\prime_0, [(v \cdot p^\prime) - \omega]^2)$.
The factor $\Delta$ is given by
$\Delta= (v \cdot p^\prime)^2 - s^\prime$.
It should be noticed the appearance of a
branch point in the perturbative spectral functions $\rho_1$ and $\rho_2$,
due to the factor $\Delta^{-{5 \over 2}}$, for small values of
$v \cdot p^\prime$, implying that the rules
cannot describe a process with small pion energy.
This is not unexpected, since the
euclidean region in the $t-$channel, where the QCD perturbative calculation
of the spectral functions can be performed,  extends
towards large
values $-q^2$, i.e.  large values of $v \cdot p^\prime$:  we shall
extrapolate the analysis of the sum rules to values of  $v \cdot p^\prime$
of the order of  $1 \; GeV$, being aware that the accuracy of the
prediction becomes poor in this range of pion energy.

The last point to be mentioned, before presenting the numerical analysis of
eqs.(\ref{eq13}, \ref{eq14}), is that
also $f_1$ and $f_2$, obtained using the effective field $h_v(x)$,
 depend on the subtraction point $\mu$;
however, in the following  we neglect this dependence since we choose to
work at the lowest order in $\alpha_s$.

In fig.1 we present the form factors $f_1$ and $f_2$ obtained from
eqs.(\ref{eq13}, \ref{eq14}) using
$s^\prime_0=0.8 \; GeV^2$, and the set of low energy parameters fixed in ref.
\cite{Neub} by the analysis of the leptonic constant
$\hat F$ in (\ref{eq12b}):
$\omega_0 = 2.1 \; GeV$, $\Lambda = 1 \; GeV$ and $\hat F=0.47 \; GeV^{3/2}$,
or
$\omega_0 = 2.5 \; GeV$, $\Lambda = 1.25 \; GeV$ and
$\hat F=0.58 \; GeV^{3/2}$. We fix the values of the Borel parameters:
$T=3 \; GeV$ and $M^{\prime 2} = 3 \; GeV^2$ after having checked the existence
of  a duality region around these values for any $v \cdot p^\prime$.

The rapid change in the behaviour of
$f_2(v \cdot p^\prime)$
for $v \cdot p^\prime \le 1.3 \; GeV$ can
be ascribed to the presence of the anomalous threshold discussed above.

Using eqs.(\ref{eq11},\ref{eq12})
 it is possible to reconstruct
$F_1(q^2)$ and $F_0(q^2)$
for intermediate values of $q^2$; the result is depicted in fig.2,
where the form factors are displayed up to $q^2=0$.
The obtained $F_1$ is
compatible with the outcome of the finite mass calculation \cite{Ball}:
it rapidly increases,
and can be fitted with the simple pole form:
$F_1(q^2) = F_1(0) / (1 - q^2 /m^2_{pole})$, with $F_1(0)=0.24$ and
$m_{pole}= 5.51 \; GeV$.
On the other hand, $F_0$ is nearly constant in $q^2$.

It is interesting to notice that the combination of $f_1$ and $f_2$ which
reconstructs $F_1$  reproduces eq.(\ref{eq8}), i.e. gives a leading
non perturbative term and a subleading perturbative contribution.
A warning is in order, however. The functions
 $f_1(v \cdot p^\prime)$ and $f_2(v \cdot p^\prime)$ obtained by
(\ref{eq13}, \ref{eq14}) are the leading terms of an expansion in
$1 /m_b$ \cite{Ligeti}; in this expansion, the pion energy
$v \cdot p^\prime$ should be kept fixed, i.e. cannot scale as $m_b$
(as it happens at $q^2=0)$ since in this case the neglected contributions
are formally
of the the same order of the terms we are considering.
On the other hand, the procedure we have followed reproduces eq.(\ref{eq8})
and is in numerical agreement with the outcome of the
calculation made at finite $m_b$ \cite{Ball}: this
suggests that the additional contributions
in the $1/m_b$ expansion
that reconstruct the form factors
should be small
 also at $q^2=0$.

Let us come back to the analytic dependence of $F_1(0)$ on $m_b$.
As mentioned above, the scaling law $1/\sqrt{m_b}$
obtained from eq.(\ref{eq8}) is
in agreement with a polar
dependence in the range from $q^2_{max}$ to $q^2=0$, with the pole given
by the $B^*$ resonance. This prediction suggests that possible
non polar components  of the form factor
near $q^2_{max}$, discussed in \cite{PR, BD1},
are extended in a limited range of $q^2$ or
they scale as $1/\sqrt{m_b}$ or faster
when $m_b \to \infty$.

Other models predict
the dependence of $F_1(0)$ on $m_b$.
In the BSW constituent quark model \cite{BSW}
 the form factors at $q^2=0$ are obtained by computing
an overlap of mesonic wave functions which are solutions of a relativistic
scalar harmonic oscillator potential;
the evolution in $q^2$ is assumed to be a simple pole. A
direct inspection  of the $m_b$ dependence of  $F_1(0)$ in this model gives
 $F_1(0) \simeq m_b^{-3/2}$, a behavior incompatible, as already
observed in \cite{BD1},
with the scaling law of $F_1(q^2_{max})$ predicted in (\ref{eq1},\ref{eq8a})
 and with
the assumption of a simple pole dependence of the form factor. The same
inconsistency is present in similar models  \cite{KS}.

The dependence of $F_1(0)$ on $m_b$ can also be predicted by light-cone QCD sum
rules \cite{Chernyak, Ruckl}. This method relies on the possibility of
computing the two-point correlator  of a quark current interpolating the
$B$-meson and of the flavour-changing current in (\ref{eq1}) calculated between
the vacuum state and the light meson state ($\pi$). Also in this case the
predicted dependence for  $F_1(0)$ is $m_b^{-3/2}$ and  therefore
the scaling law is incompatible with the correct behaviour at
$q^2_{max}$ and with the claim \cite{Ruckl} that the observed $q^2$ dependence
of the form factor
is polar. The origin of such inconsistency is currently under investigation.

Let us conclude with a comment on the results of the present study.
The measurement of $V_{ub}$, which is of prime importance in the phenomenology
of the Standard Model, cannot be performed by the simple observation of the
decay $B \to \pi \ell \nu$ without referring to models for the evaluation of
the hadronic matrix element.
Here we have obtained indications that $F_1$ is polar in $q^2$, and that
the scaling law $F_1(0) \simeq 1/\sqrt{m_b}$ should be fulfilled, at least
asymptotically in the heavy quark mass. This information allows to
use the full sample of pions from semileptonic $B$ decays; $V_{ub}$ could be
determined by
comparing eq.(\ref{eq4}) with the analogous expression
for $D \to \pi \ell \nu$ and using the experimental result of this last decay.
The only drawback, as usual in these considerations, is represented by the
size of $1/m_Q$ corrections, an argument which deserves an independent
investigation.

\vskip 1.5cm
\noindent {\bf Acknowledgments}
\vskip 0.5cm
\noindent
We thank P.Ball, F.Buccella, G.Nardulli and N.Paver for fruitful discussions.

\newpage

{\bf\Large
\centerline{Figure Captions}}

\vspace{2truecm}

{\bf Fig 1:} The form factors $f_1(v\cdot p')$ and $f_2(v\cdot p')$.

\vspace{1truecm}

{\bf Fig 2:} The form factors $F_1(q^2)$ and $F_0(q^2)$.

\end{document}